\documentclass[epj]{webofc}
\usepackage[utf8]{inputenc}
\usepackage[varg]{txfonts}   
\usepackage{booktabs}
\usepackage{xcolor}
\definecolor{darkred}{rgb}{0.4,0.0,0.0}
\definecolor{darkgreen}{rgb}{0.0,0.4,0.0}
\definecolor{darkblue}{rgb}{0.0,0.0,0.4}
\usepackage[bookmarks,linktocpage,colorlinks,
    linkcolor = darkred,
    urlcolor  = darkblue,
    citecolor = darkgreen]{hyperref}
\usepackage{subfigure}
\wocname{EPJ Web of Conferences}
\woctitle{Lattice2017}
\begin{document}
%
\selectlanguage{english}
\title{%
SU(2) with fundamental fermions and scalars
}
\author{%
\firstname{Martin} \lastname{Hansen}\inst{1}  \and
\firstname{Tadeusz} \lastname{Janowski}\inst{1} \and
\firstname{Claudio} \lastname{Pica}\inst{1} \and
\firstname{Arianna}  \lastname{Toniato}\inst{1}\fnsep\thanks{Speaker, \email{toniato@cp3.sdu.dk}}
}
\institute{%
CP$^3$-Origins, University of Southern Denmark, Campusvej 55, DK-5230 Odense M, Denmark
}
\abstract{%
 We present preliminary results on the lattice simulation of an SU(2) gauge theory with two fermion flavors and one strongly interacting scalar field, all in the fundamental representation of SU(2). The motivation for this study comes from the recent proposal of ``fundamental'' partial compositeness models featuring strongly interacting scalar fields in addition to fermions. Here we describe the lattice setup for our study of this class of models and a first exploration of the lattice phase diagram. In particular we then investigate how the presence of a strongly coupled scalar field affects the properties of light meson resonances previously obtained for the SU(2) model.
 \\[.1cm]
 {\footnotesize  \it Preprint: CP3-Origins-2017-047 DNRF90  }
}
\maketitle

\section{Introduction}\label{intro}

Composite Higgs models \cite{Weinberg:1975gm,Susskind:1978ms,Kaplan:1983fs,Kaplan:1983sm} aim to provide an alternative to the Standard Model's Higgs mechanism for Electroweak symmetry breaking and mass generation. These alternative mechanisms are based on the introduction of a new strongly interacting sector and they can be broadly divided into Technicolor and composite Goldstone Higgs models. In Technicolor models~\cite{Weinberg:1975gm,Susskind:1978ms}, the $W^{\pm}$ and $Z$ masses are generated as a consequence of chiral symmetry breaking in the new strong sector: specifically three of the Goldstone bosons provide the longitudinal degrees of freedom of the $W^{\pm}$ and $Z$ gauge bosons. In Technicolor models the Higgs boson is the lightest scalar resonance of the new strongly interacting sector.
In composite Goldstone Higgs models~\cite{Kaplan:1983fs,Kaplan:1983sm}, an enlarged global flavor symmetry of the new strong sector breaks in such a way to preserve the electroweak symmetry and the Higgs boson is identified with one of the additional Goldstone bosons. In order to break electroweak symmetry and generate masses, additional interactions are introduced which have the effect of triggering electroweak symmetry breaking, generating the right mass for the composite Higgs, and possibly generate the correct mass for other SM particles. These models have the advantage of featuring a naturally light Higgs, with a large separation between the electroweak scale and the scale at which the fermion condensate occurs at the expense of a moderate tuning.

A delicate point within the context of a composite Higgs model is the generation of the Standard Model fermion masses. This is usually done by introducing four-fermion operators which couple the fermions of the new strong sector and the ordinary Standard Model fermions. 
These four-fermion operators are non-renormalizable, and have to be thought of as effective operators, for example as arising once some heavy mediator has been integrated out. 
There are two popular paradigms to generate the Standard Model fermion masses: extended Technicolor~\cite{Eichten:1979ah}, and partial compositeness~\cite{Kaplan:1991dc}. 
In extended Technicolor, couplings between two Standard Model fermions ($f$) and two techni-fermions from the new strong sector ($\mathcal{F}$)  are introduced, while in partial compositeness the four fermion operators couple one Standard Model fermion with a composite fermionic operator made of three (hyper-)fermions from the new sector. 
We can schematically write those operators as $(\lambda/\Lambda_{UV}^2) \bar{f} f \bar{\mathcal{F}} \mathcal{F}$ in the case of extended Technicolor, and $(\lambda/\Lambda_{UV}^2) \bar{f} \mathcal{F} \mathcal{F} \mathcal{F}$ in the case of partial compositeness, where $\Lambda_{UV}$ represents the mass of the heavy mediator that has been integrated out, and $\lambda$ a dimensionless coupling constant.
Once chiral symmetry is broken in the new strong sector, those operators provide masses for the Standard Model fermions.

Both mechanisms encounter severe difficulties in generating large masses, such as for example the top quark mass. In fact in order to respect the experimental bounds on flavor changing neutral currents the scale $\Lambda_{UV}$ must be high ($ \gtrsim 1000~\mathrm{TeV}$), which suppresses the values of the fermion masses. 
Near conformal dynamics with a ``walking'' gauge coupling, together with large anomalous dimensions for the composite operators $\bar{\mathcal{F}} \mathcal{F}$ or $\mathcal{F} \mathcal{F} \mathcal{F}$, can solve the problem by removing the large suppression given by the running from $\Lambda_{UV}$ to the energy scale at which the fermion condensate occurs in the new strong sector~\cite{Holdom:1981rm,Holdom:1984sk,Akiba:1985rr,Appelquist:1986an,Yamawaki:1985zg,Appelquist:1986tr,Appelquist:1987fc}. 
In particular, in the case of partial compositeness the $\mathcal{F} \mathcal{F} \mathcal{F}$ operator is required to have an anomalous dimension $\gamma \approx 2$, corresponding to a scaling dimension of $3 \times 3/2 - \gamma \approx 5/2$. 
However it seems that such a large anomalous dimension is hard to achieve. 
Recent studies of SU(3) gauge theories with fundamental fermions inside the conformal window indicate that for these models the anomalous dimension of baryonic operators remains small down to the lowest point in the conformal window that can be reached within perturbation theory~\cite{Pica:2016rmv}. This challenges the minimal models of partial compositeness in which baryonic operators composed by three fermions of the new sector are coupled to Standard Model fermions. 

In order to overcome these difficulties, a model of "fundamental" partial compositeness has recently been proposed~\cite{Sannino:2016sfx}. 
This is a composite Goldstone Higgs model in which the new strongly interacting sector contains as matter fields not only fermions ($\mathcal{F}$), but also strongly interacting scalars ($\mathcal{S}$). 
The partial compositeness mechanism is implemented by introducing operators of the form $\lambda \bar{f} \mathcal{F} \mathcal{S}$, which couple one Standard Model fermion with one baryonic operator of the new strong sector. 
The baryonic operator $\mathcal{F} \mathcal{S}$ has an engineering dimension of $5/2$, therefore requiring no large anomalous dimensions. 
It can also be argued that since any purely fermionic model of partial compositeness must involve baryons with scaling dimensions close to $5/2$, these baryons would effectively behave as if they were made by a fermion and a composite scalar similar to the fundamental scalar appearing in this model~\cite{Sannino:2016sfx}.
An interesting feature of this new kind of composite Higgs model is that it can be used to simultaneously explain the $R_K$, $R_{K^*}$ and $g-2$ anomalies provided the Yukawa coupling to the muon is large ($\sim 1.5$)~\cite{DAmico:2017mtc}. 
A minimal model of fundamental partial compositeness able to generate masses for all the Standard Model fermions with 2 fermions and 12 scalars in the new strong sector is currently under investigation ~\cite{Cacciapaglia:2017cdi}.

An important assumption underlying the construction of such fundamental partial compositeness models is that the presence of strongly interacting scalars does not significantly modify the dynamics of the new strong sector. 
In particular the symmetry breaking pattern must remain the same as if only fermions are present.
Lattice simulations are required in order to investigate how scalar fields affect the strong dynamics of a gauge-fermion theory. With these motivations we have recently started the study of an SU(2) gauge theory with fundamental fermions and scalars on the lattice. 
Lattice simulations do not intend to simulate a whole partial compositeness model, including all couplings between the new strong sector and the Standard Model, but they aim to analyse the new strong sector in isolation. Here, for our first exploratory study we simulate an SU(2) gauge theory with two fundamental fermions and one fundamental colored scalar.

This proceeding is organised as follows: in section \ref{couplings} we discuss some relevant features of the running of the scalar quartic couplings for continuum theories, in section~\ref{setup} we describe the lattice setup of our simulations, in section~\ref{results} we show our preliminary results on the meson spectrum and phase structure of the model and in section~\ref{conclusions} we briefly conclude and outline future directions of work.

\section{Running of the scalar quartic couplings}\label{couplings}

We start our analysis by discussing some relevant and interesting features of the RG running of scalar quartic couplings for continuum theories. Quartic couplings among strongly interacting scalars are always generated at loop level, e.g. via gauge interactions. While generically one expects these quartic couplings to present a UV Landau pole, one interesting feature is that this is not necessarily the case. 
In fact these models can be completely asympotically free, i.e. all couplings will vanish at very high-energies. As a simple example here we consider an SU(N) gauge theory with $N_f$ fundamental fermions and $N_S$ fundamental scalars. The most general scalar quartic potential, containing only SU($N_S$) flavor-symmetric operators is given by: 

\begin{equation}
V = \lambda^{(1)} [ \mathrm{Tr} (\mathcal{S}^{\dagger} \mathcal{S}) ]^2 + \lambda^{(2)} \mathrm{Tr} (\mathcal{S}^{\dagger} \mathcal{S} \mathcal{S}^{\dagger} 
 \mathcal{S})
 \label{potential}
\end{equation}
where $\mathcal{S}$ is a complex $N \times N_S$ matrix, carrying a color and a flavor index. If $N_S = 1$ then the two operators in equation~\ref{potential} are equal: $[ \mathrm{Tr} (\mathcal{S}^{\dagger} \mathcal{S}) ]^2 = \mathrm{Tr} (\mathcal{S}^{\dagger} \mathcal{S} \mathcal{S}^{\dagger} \mathcal{S})$. In this case there is one single scalar quartic coupling $\lambda =  \lambda^{(1)} + \lambda^{(2)}$. For sake of simplicity we restrict ourselves to the case of $N_S = 1$, a more complete analysis can be found in~\cite{Hansen:2017pwe}.

\begin{figure}[thb] 
  \centering
  \includegraphics[width=6.4cm,clip]{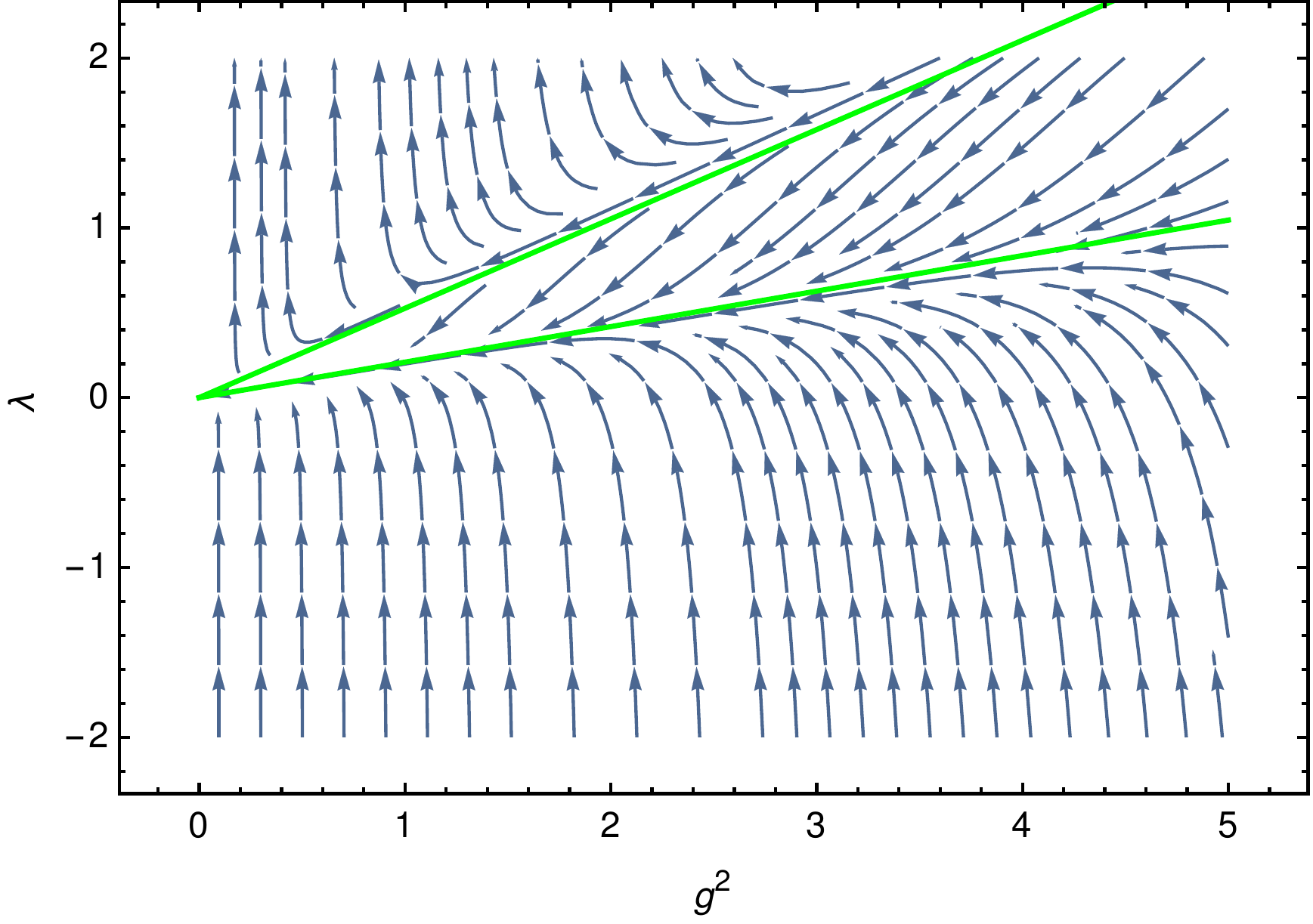} ~~~~~~~~~~~
  \includegraphics[width=6.5cm,clip]{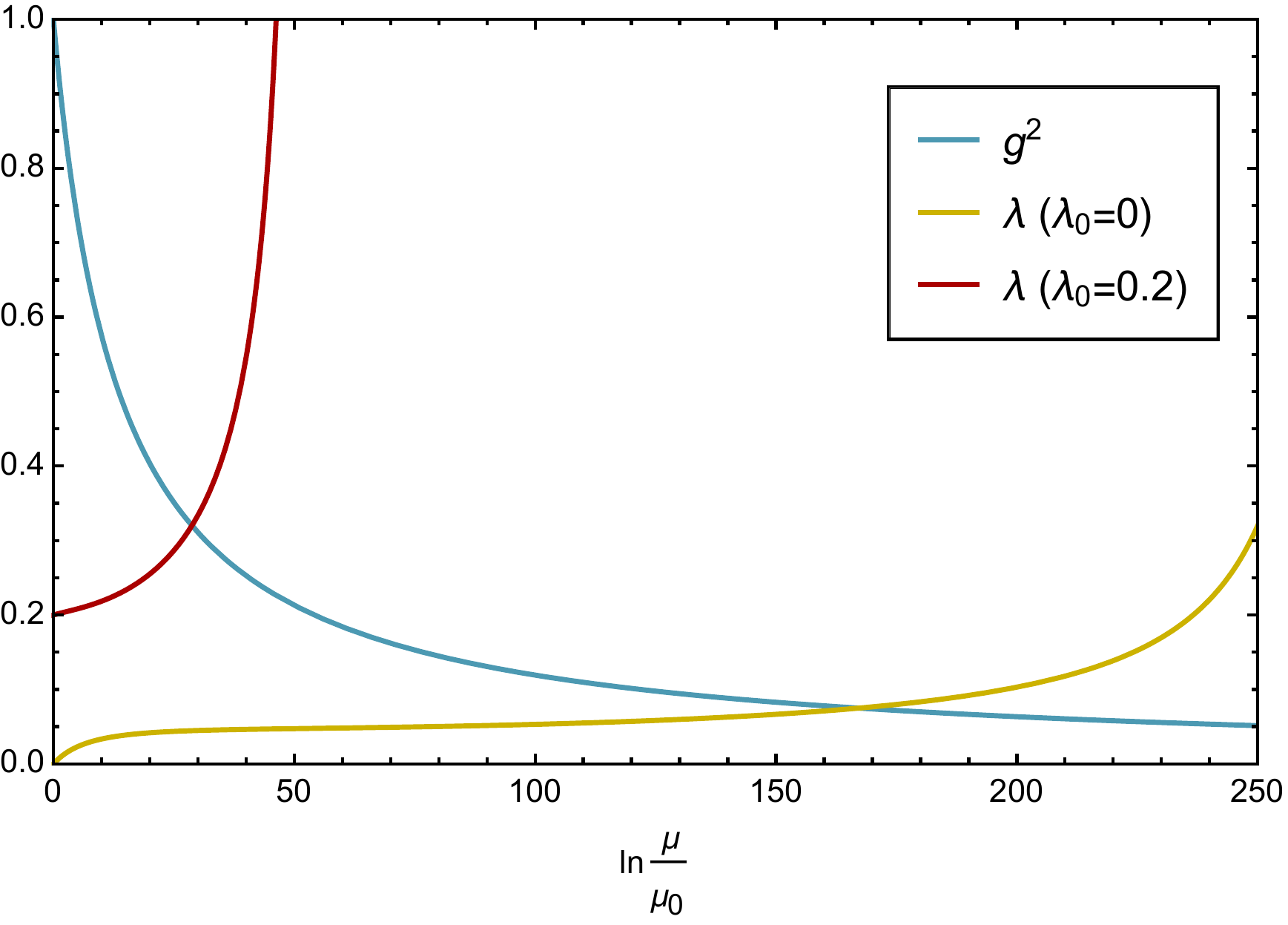}
  \caption{Left panel: Running of the squared gauge coupling $g^2$ and of the scalar quartic coupling $\lambda$ for a theory with SU(5) gauge group, $N_f=26$ fundamental fermions and $N_S =1$ fundamental scalars. The arrows indicate the direction of increasing energy. The green lines represent the fixed flow solutions of the RG equations, characterised by a constant ratio $\lambda/g^2$.
 Right panel: Running of $g^2$ and $\lambda$ as functions of $\ln (\mu/\mu_0)$ in an SU(2) gauge theory with $N_f=2$ fundamental fermions and $N_S=1$ fundamental scalars. $\lambda$ is plotted for two different initial conditions: $\lambda_0 = \lambda(\mu_0) = 0$, and $\lambda_0=\lambda(\mu_0)=0.2$.}
  \label{flow}
\end{figure}

The one-loop $\beta$-functions for the gauge coupling $g$ and the scalar quartic coupling $\lambda$ are given by: 

\begin{equation}
\begin{split}
 & (4 \pi)^2 \beta_g = - \biggl( \frac{11}{3} N - \frac{2}{3} N_f - \frac{N_S}{6} \biggr) g^3 \, ,\\
 & (4 \pi)^2 \beta_{\lambda} = 4(N +4) \lambda^2 -\frac{6(N^2 -1)}{N} g^2 \lambda + \frac{3N^3 + 3N^2 -12N + 6}{4 N^2} g^4 \, ,
\end{split}
\end{equation}
where the first equation is general and holds for any $N_S$, while the second one holds only for $N_S=1$. The solutions of the renormalization group equations $ \mathrm{d} \alpha / \mathrm{d} \ln \mu = \beta_{\alpha}$  ($\alpha = g, \lambda$) indicate that $N$ and $N_f$ can be chosen in such a way to have complete asymptotic freedom, which means that both $g$ and $\lambda$ flow to zero in the UV (large $\mu$), and this phenomenon normally occurs when a large number of fermions is present. As an example of this behavior, the left panel of figure~\ref{flow} shows the running of $g^2$ and $\lambda$ in the case of $N=5$ and $N_f=26$, where the arrows point in the direction of increasing energy. It can be seen that, while $g^2$ always goes to zero at high energies, the initial condition for $\lambda$ can be chosen in such a way that also $\lambda$ flows to zero in the UV, realising complete asymptotic freedom. 

In table \ref{asympt_freedom} we report, for different values of $N$, the ranges in $N_f$ such that there exist completely asymptotically free solutions. 
It can be noticed that for $N=2$, which is the object of our lattice study, there is no possible choice of $N_f$ leading to complete asymptotic freedom. Nevertheless the case $N=2$ is of great interest, because the SU(2) model with $N_f=2$ fundamental fermions has been extensively studied~\cite{Arthur:2016dir}, and those studies offer a benchmark for seeing how adding a strongly interacting scalar modifies the dynamics of the model. The right panel of figure~\ref{flow} shows the running of $g^2$ and $\lambda$ in the theory that we simulate on the lattice: SU(2) with $N_f=2$ fundamental fermions and $N_S=1$ fundamental scalars. 
Two different initial conditions are chosen for $\lambda$: $\lambda_0 = \lambda(\mu_0) = 0$, and $\lambda_0=\lambda(\mu_0)=0.2$. 
While in both cases UV Landau poles for $\lambda$ are present, these occur at an energy scale which is very large compared to the scale at which $g^2=1$. Specifically, if we assume the scale at which $g^2=1$ to be $\mu_0 \sim 10^3 \mathrm{GeV}$, then the Landau pole is located beyond the Planck scale ($\ln(\mu_{Planck} / \mu_0) \sim 37$) for both choices of initial conditions. 
This means that the continuum theory is a good effective theory at the scale at which the interesting electroweak physics happens.

For the lattice model, the absence of a UV fixed point for $\lambda$, from the point of view of Wilsonian renormalization, makes it impossible to take a continuum limit. 
However given the large separation of scales between the interesting physics and the Landau pole energy, we expect that the lattice model of this theory will present a good scaling window when changing the lattice spacing, which can therefore provide valuable input for phenomenological models. 

\begin{table}[thb]
  \small
  \centering
  \caption{For different values of N, ranges in $N_f$ such that there exist completely asymptotically free solutions, in a theory with SU(N) gauge group, $N_f$ fundamental fermions and $N_S=1$ fundamental scalars.}
  \begin{tabular}{lll}\toprule
  $N$ & $N_f$  \\\midrule
  2 & No solutions \\
  3 & $15.93 < N_f < 16.25$ \\ 
  4 & 1$9.8 < N_f < 21.75$ \\
  5 & $23.56 < N_f < 27.25$ \\
  6 & $27.27 < N_f  < 32.75$ \\
  7 & $30.94 < N_f  < 38.25$ \\
  8 & $34.6 < N_f  < 43.75$ \\
  9 & $38.24 < N_f < 49.25$ \\
  10 & $41.88 < N_f < 54.75$ \\\bottomrule
  \end{tabular}
  \label{asympt_freedom}
\end{table}


\section{SU(2) + 2 fermions + 1 scalar: lattice setup}\label{setup}
\begin{samepage}
We simulate a theory with SU(2) gauge group, $N_f=2$ fundamental fermions and $N_S=1$ fundamental scalars. Our lattice action is given by the sum of three contributions: $\mathcal{S} = \mathcal{S}_G + \mathcal{S}_F + \mathcal{S}_S$, with $\mathcal{S}_G$ the Wilson plaquette action, $\mathcal{S}_F$ the contribution of two mass-degenarate Wilson fermions, and $\mathcal{S}_S$ the scalar contribution given by:
\end{samepage}
\begin{equation}
\mathcal{S}_S =  \sum_x \biggl[ - \sum_{\mu} \biggl( S^{\dagger}(x) U(x, \mu) S(x+\mu) + S^{\dagger}(x) U(x-\mu,\mu)^{\dagger} S(x-\mu) \biggr) +  
 (m_S^2+8) S^{\dagger}(x) S(x) + \lambda (S^{\dagger}(x) S(x))^2 \biggr]
\end{equation}
where $U_{\mu}(x)$ are the gauge link variables and $S(x)$ is the scalar field, in the fundamental representation of SU(2). 

There are four bare parameters: the inverse lattice gauge coupling $\beta$, the mass of the fermion fields $m_f$, the mass of the scalar field $m_S$ and the scalar quartic coupling $\lambda$. For our first exploratory analysis we choose one single lattice volume ($T=32$, $V=16^3$) and we fix $\beta=2.0$ and $m_f=-0.94$, based on our previous simulations without the scalar field. We simulate two different values of the scalar quartic coupling: $\lambda=0$, $\lambda=2$, and different values of the squared scalar mass which are reported in table~\ref{parameters}. We simulate both positive and negative values of the squared scalar mass, in order to try to identify the region in the parameter space where symmetry breaking occurs in the scalar sector.

We generate the configurations of the gauge and scalar fields using the HiRep code~\cite{DelDebbio:2008zf}, which we extended in order to simulate the scalar field along with gauge and fermions. In particular the scalar field is treated as a dynamical field, with associated conjugate momentum, in the the HMC algorithm.

\begin{table}[thb]
 \small
  \centering
  \caption{Values of the scalar quartic coupling $\lambda$ and the squared scalar mass $m_S^2$ used in our simulations.}
          \begin{tabular}{lrrrrrrrrr}\toprule
  $\lambda$ & $m_S^2$ \\\midrule
  0 & 25 & 6.25 & 1.00 & 0.25 & 0.00 & -0.25 & -0.50 & -0.75 & -1.00 \\
  2 & 25 & 6.25 & 1.00 & 0.25 & -0.50 & -1.00 & -1.20 & -2.00 & -3.00  \\\bottomrule
  \end{tabular}
      \label{parameters}
\end{table}

\begin{figure}[thb] 
  \centering
  \includegraphics[width=7.05cm,clip]{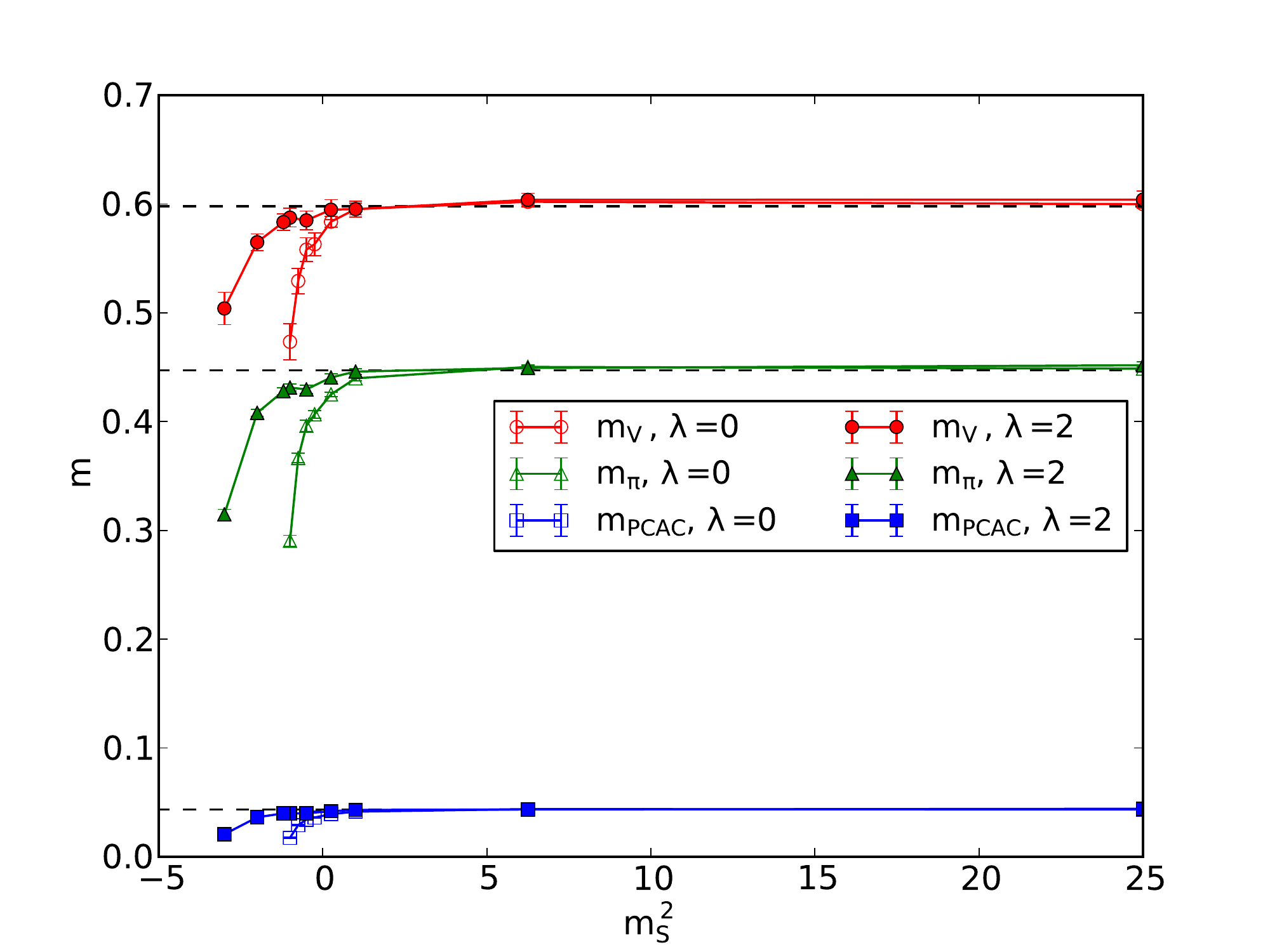} 
  \includegraphics[width=7.05cm,clip]{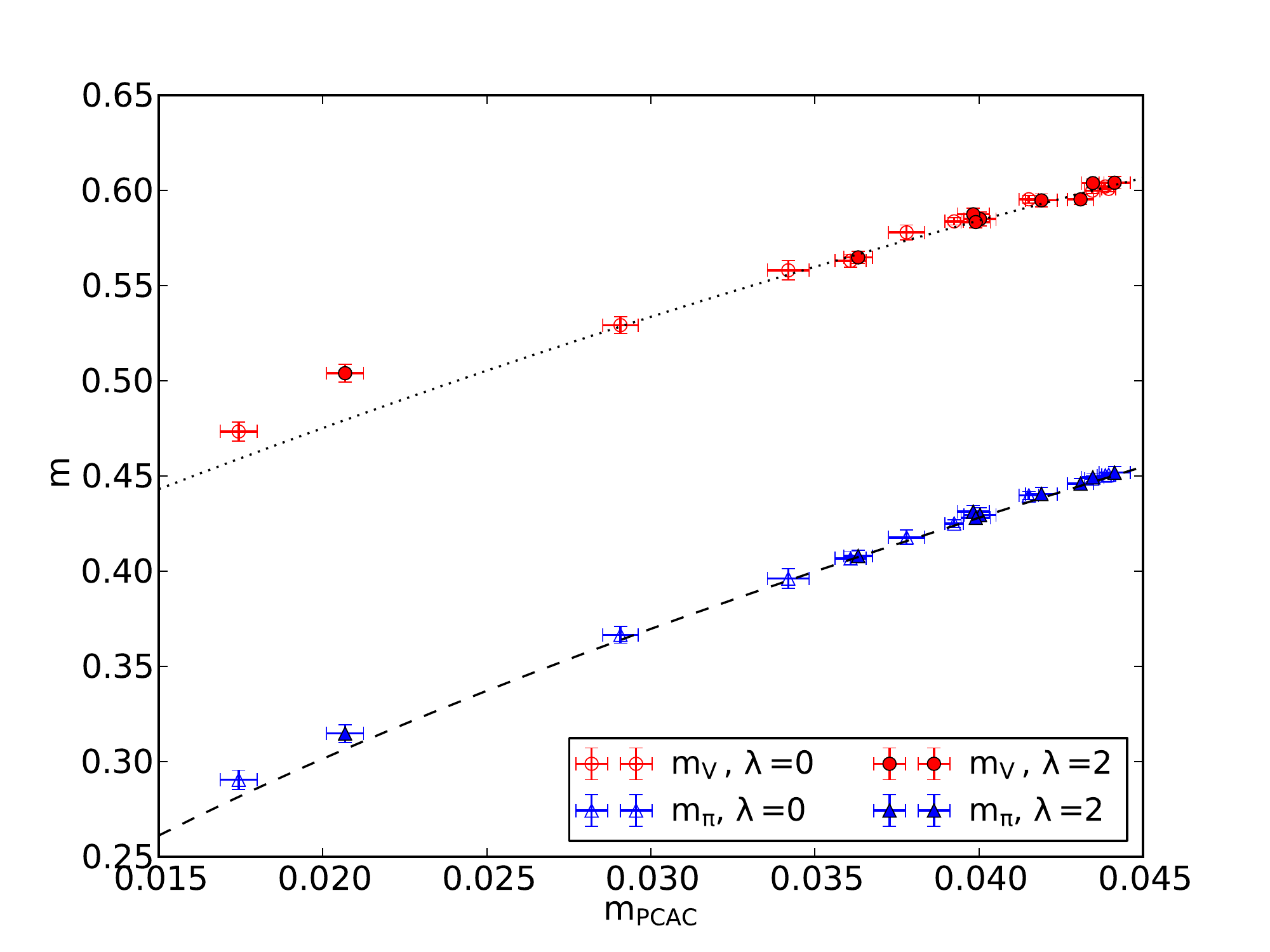}
  \caption{Left panel: Bare mass of the pseudoscalar meson ($m_{\pi}$), of the vector meson ($m_V$) and bare PCAC mass ($m_{PCAC}$) as functions of the squared scalar mass $m_S^2$ for two different values of $\lambda$, together with the results for the same observables obtained in the gauge-fermion model (dashed lines). 
 Right panel: $m_{\pi}$ and $m_V$ as functions of $m_{PCAC}$. The dashed line represents a chiral fit of $m_{\pi}$ as a function of $m_{PCAC}$ obtained in the gauge-fermion model, and the dotted line a polynomial fit of $m_V$ as a function of $m_{PCAC}$, always obtained from the data of the gauge-fermion model.}
  \label{mesons}
\end{figure}

\begin{nobreak}
\section{Preliminary results}\label{results}

We start our exploratory analysis by concentrating on the spectrum of light meson resonances, and on a first scan of the parameter space. In particular we focus on how the presence of the scalar field modifies the meson masses which have been previously calculated for the gauge-fermion model~\cite{Arthur:2016dir}, and for each choice of the parameters $\lambda$ and $m_S^2$ we examine the structure of the effective potential of the scalar field, in order to understand  if symmetry breaking occurs in the scalar sector.

\subsection{Meson spectrum}

We measure the spectrum of light meson resonances at fixed $\beta=2.0$ and $m_f = -0.94$ and for the values of $\lambda$ and $m_S^2$ reported in table~\ref{parameters}. The results are shown in figure~\ref{mesons}. In the left panel the bare mass of the pseudoscalar meson ($m_{\pi}$), of the vector meson ($m_V$) and the bare PCAC mass are plotted as functions of the squared scalar mass $m_S^2$, together with the results for the same observables measured in the gauge-fermion model~\cite{Arthur:2016dir} (dashed lines). It can be observed that the results of the gauge-fermion model are reproduced for $m_S^2 \gtrsim 1$, below this value the presence of the scalar starts to modify the meson and PCAC masses.

In the right panel of figure~\ref{mesons}, $m_{\pi}$ and $m_V$ are plotted as functions of $m_{PCAC}$, together with fits for the same observables obtained in the gauge-fermion model. In this plot the dashed line represents a chiral fit of $m_{\pi}$ as a function of $m_{PCAC}$, and the dotted line represents a polynomial fit of $m_V$ as a function of $m_{PCAC}$, both obtained from the data of the gauge-fermion simulations without scalar field~\cite{Arthur:2016dir}. 
This plot shows that even though the presence of the scalar field significantly changes the values of the meson masses and of the PCAC mass, the relationship between the meson masses and $m_{PCAC}$ remains largely unchanged.
\end{nobreak}

\subsection{Phase structure}

Our first analysis of the phase structure of the model is meant to identify the region in the parameter space where symmetry breaking occurs in the scalar sector. In order to do so, we examine the structure of the effective potential of the scalar field. 

 A method to compute the effective potential of a scalar field has been proposed and applied to the real scalar lattice field theory in \cite{Kuti:1987bs}. Here we apply the same method to our complex SU(2) doublet scalar, in a specific gauge described below. It is pointed out in \cite{Kuti:1987bs} that the constraint effective potential $U_{\Omega}(\bar{S})$ \cite{ORaifeartaigh:1986axd,Fukuda:1974ey} is related to the probability $P(\bar{S})$ to find the system in a state characterised by $\bar{S} = \Omega^{-1} \sum_x S(x)$ by the following relation:
 
 \begin{equation}
 P(\bar{S}) = \frac{e^{-\Omega U_{\Omega}(\bar{S})}}{\int d\bar{S} e^{-\Omega U_{\Omega}(\bar{S})}}
 \end{equation}
 where $\Omega$ represents the finite lattice volume. The values of $\bar{S}$ that maximise the probability $P(\bar{S})$ are the minima of the effective potential $U_{\Omega}(\bar{S})$.
 
 We proceed as follows: first we use a gauge transformation to express the scalar field in the following form

  \begin{align}
  S(x) & = 
    \begin{pmatrix}
     S_1(x) \\
     S_2(x)
    \end{pmatrix}  \to \tilde{S}(x) = \mathrm{sign}(\mathrm{Re}(S_1(x)))
    \begin{pmatrix}
    \sqrt{S^{\dagger}(x) S(x)} \\
    0
    \end{pmatrix} \:,
 \end{align}
and then we construct histograms of the distribution of $\bar{S} = \Omega^{-1} \sum_x \tilde{S}(x)$ among the configurations. In this setup we expect to observe a double-peak structure in the distribution (corresponding to a double-well potential) in the broken-symmetry phase, and a single-peak structure (single-well potential) in the symmetric phase.

\begin{figure}[thb] 
  \centering
  \includegraphics[width=16cm,clip]{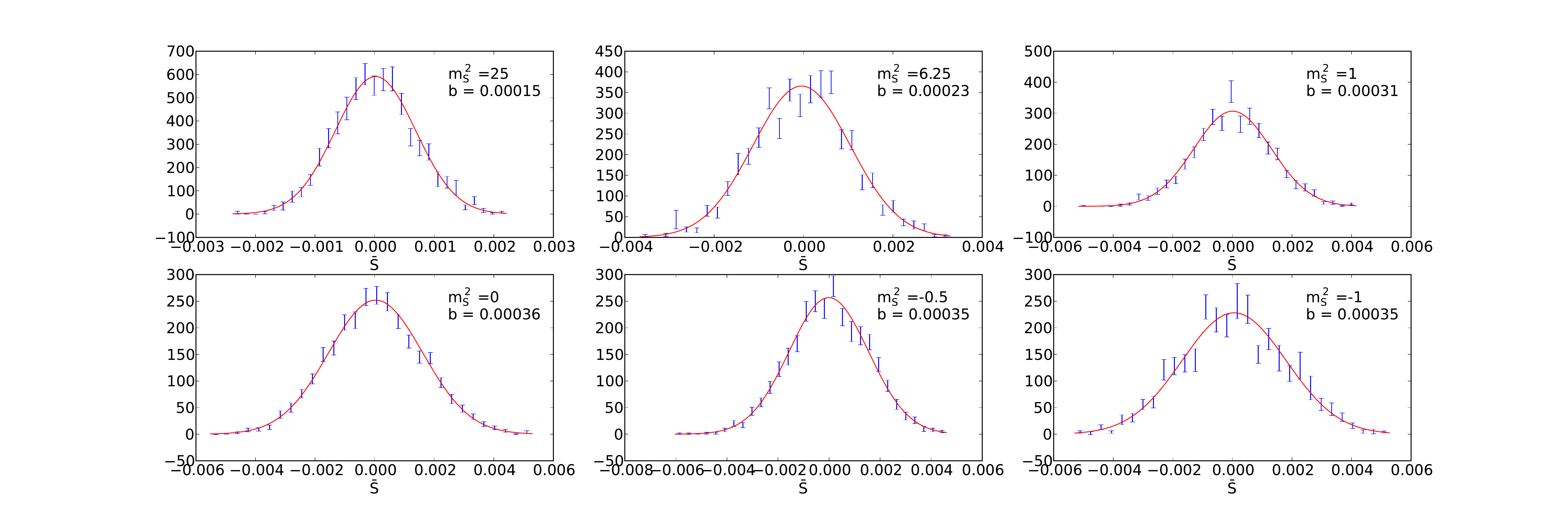} 
  \caption{Histograms of the distribution of $\bar{S}$ among the configurations, for $\lambda=0$ and different values of $m_S^2$, together with gaussian curves calculated with the average value and standard deviation obtained from the $\bar{S}$ sample. The bin size $b$ is reported together with each histogram.}
  \label{histograms}
\end{figure}

 In the range of parameters studied we always observed a single-peak structure. As an example we report in figure~\ref{histograms} the histograms that we obtained for $\lambda=0$ and different values of $m_S^2$. From the results of this first analysis it seems that we did not observe symmetry breaking in the scalar sector in the range explored so far.

%

\begin{samepage}
\section{Conclusions}\label{conclusions}

Motivated by the recent proposal of composite Higgs models featuring strongly interacting scalars in addition to fermions~\cite{Sannino:2016sfx}, we started lattice simulations of an SU(2) gauge theory with two fermions and one scalar field in the fundamental representation of the gauge group. We presented here our first results on the meson spectrum and a preliminary analysis of the phase structure of the model.

After fixing the inverse gauge coupling $\beta$ and the fermion mass $m_f$ to values which have already been used in the SU(2) gauge theory with two fundamental fermions~\cite{Arthur:2016dir}, we simulated the model for several values of the squared scalar mass $m_S^2$ and scalar quartic coupling $\lambda$. We observe that, provided $m_S^2$ is large enough, the meson spectrum of the gauge-fermion model is recovered. For smaller $m_S^2$ large deviations can be observed, which however do not seem to affect the dependence of the meson masses on the PCAC mass. 

In order to localise the phase in which spontaneous symmetry breaking occurs in the scalar sector, we simulated both positive and negative values of $m_S^2$ and we analysed the structure of the effective potential of the scalar field in a fixed gauge. In the range of parameters studied, we have not observed symmetry breaking in the scalar sector. 
\nopagebreak
In our future analysis we plan to extend the current analysis to more values of $\beta$, $m_f$, $m_S^2$ and $\lambda$ and use different lattice volumes, in order to precisely outline the phase structure of the model. We also plan to investigate the spectrum of bound states composed by two scalars and by one fermion and one scalar relevant for the partial compositeness scenario.

\section*{Acknowledgements}
This work was supported by the Danish National Research Foundation DNRF:90 grant and by a Lundbeck Foundation Fellowship grant. The computational resources were provided by the DeIC national HPC centre at SDU.
\end{samepage}
\clearpage
\bibliography{lattice2017}

\begin{thebibliography}{23}

\bibitem{Weinberg:1975gm}
S.~Weinberg, Phys. Rev. \textbf{D13}, 974 (1976)

\bibitem{Susskind:1978ms}
L.~Susskind, Phys. Rev. \textbf{D20}, 2619 (1979)

\bibitem{Kaplan:1983fs}
D.B. Kaplan, H.~Georgi, Phys. Lett. \textbf{136B}, 183 (1984)

\bibitem{Kaplan:1983sm}
D.B. Kaplan, H.~Georgi, S.~Dimopoulos, Phys. Lett. \textbf{136B}, 187 (1984)

\bibitem{Eichten:1979ah}
E.~Eichten, K.D. Lane, Phys. Lett. \textbf{B90}, 125 (1980)

\bibitem{Kaplan:1991dc}
D.B. Kaplan, Nucl. Phys. \textbf{B365}, 259 (1991)

\bibitem{Holdom:1981rm}
B.~Holdom, Phys. Rev. \textbf{D24}, 1441 (1981)

\bibitem{Holdom:1984sk}
B.~Holdom, Phys. Lett. \textbf{150B}, 301 (1985)

\bibitem{Akiba:1985rr}
T.~Akiba, T.~Yanagida, Phys. Lett. \textbf{169B}, 432 (1986)

\bibitem{Appelquist:1986an}
T.W. Appelquist, D.~Karabali, L.C.R. Wijewardhana, Phys. Rev. Lett.
  \textbf{57}, 957 (1986)

\bibitem{Yamawaki:1985zg}
K.~Yamawaki, M.~Bando, K.i. Matumoto, Phys. Rev. Lett. \textbf{56}, 1335 (1986)

\bibitem{Appelquist:1986tr}
T.~Appelquist, L.C.R. Wijewardhana, Phys. Rev. \textbf{D35}, 774 (1987)

\bibitem{Appelquist:1987fc}
T.~Appelquist, L.C.R. Wijewardhana, Phys. Rev. \textbf{D36}, 568 (1987)

\bibitem{Pica:2016rmv}
C.~Pica, F.~Sannino, Phys. Rev. \textbf{D94}, 071702 (2016)

\bibitem{Sannino:2016sfx}
F.~Sannino, A.~Strumia, A.~Tesi, E.~Vigiani, JHEP \textbf{11}, 029 (2016)

\bibitem{DAmico:2017mtc}
G.~D'Amico, M.~Nardecchia, P.~Panci, F.~Sannino, A.~Strumia, R.~Torre,
  A.~Urbano, JHEP \textbf{09}, 010 (2017)

\bibitem{Cacciapaglia:2017cdi}
G.~Cacciapaglia, H.~Gertov, F.~Sannino, A.E. Thomsen (2017),
  \texttt{1704.07845}

\bibitem{Hansen:2017pwe}
F.F. Hansen, T.~Janowski, K.~Langaeble, R.B. Mann, F.~Sannino, T.G. Steele,
  Z.W. Wang (2017), \texttt{1706.06402}

\bibitem{Arthur:2016dir}
R.~Arthur, V.~Drach, M.~Hansen, A.~Hietanen, C.~Pica, F.~Sannino, Phys. Rev.
  \textbf{D94}, 094507 (2016)

\bibitem{DelDebbio:2008zf}
L.~Del~Debbio, A.~Patella, C.~Pica, Phys. Rev. \textbf{D81}, 094503 (2010)

\bibitem{Kuti:1987bs}
J.~Kuti, Y.~Shen, Phys. Rev. Lett. \textbf{60}, 85 (1988)

\bibitem{ORaifeartaigh:1986axd}
L.~O'Raifeartaigh, A.~Wipf, H.~Yoneyama, Nucl. Phys. \textbf{B271}, 653 (1986)

\bibitem{Fukuda:1974ey}
R.~Fukuda, E.~Kyriakopoulos, Nucl. Phys. \textbf{B85}, 354 (1975)

\end{thebibliography}

\end{document}